# A Low-Profile Polarization-Adjustable Ring Slot Antenna for Millimeter-wave Applications

P. Aghabeyki, M. Hamidi, G. Moradi, and P. Mousavi

*Abstract*—In this letter, a dual port antenna with polarization adjustability is proposed. The loading effect of four ring slots stimulates a hybrid mode in the antenna that generates eight magnetic monopoles with high gain in boresight. Due to symmetrical configuration, adding a 90°-rotated port provides a dual polarized antenna. Through controlling the phase difference between the two ports, six types of polarization can be demonstrated. A prototype is fabricated and measured in 28 GHz. Antenna shows a maximum gain of 13 dBi, X% bandwidth and a good cross-polarization level in all polarization states. Distinct features including low-profile, low-cost, single-layered and no feeding network makes the antenna a suitable candidate for upcoming millimeter-wave applications.

## I. Introduction

POLARIZATION flexibility has always been an exciting feature for various applications. By the advent of 5G and other millimeter-wave technologies, this feature could be a promising solution to increase the channel capacity, alleviate the multipath fading effect and reduce the size of the system. Moreover, antennas with dual polarization characteristics attracted significant attention in MIMO applications, especially for the base station. Therefore, developing an antenna with these properties can be very beneficial [1].

Numerous studies have been done in low frequencies to realize reconfigurable antennas with integrated switching system [2]-[5]. However, in millimeter-wave frequencies, integrating an active circuit is challenging; hence limited studies focused on this capability. Micro-electromechanical system (MEMS) were employed in [6] and [7] to realize a reconfigurable polarization antenna at 77 GHz and 60 GHz respectively.

Due to these challenges, A few studies have been done to develop adjustable polarization antennas for millimeter-wave applications [8]-[12]. In [8], a 3-layer aperture-coupled microstrip antenna is presented in order to realize six types of polarizations. In [9], a microstrip antenna array is proposed at 60 GHz. The antenna has two input ports to be adjusted with a proper phase difference, and a feeding network comprises branch line couplers and a power divider. In [10], a 4-port antenna with quadri-polarization is reported at 60 GHz. The antenna is realized on three substrate layers, two of which designated for feeding network and one for radiating elements. A microstrip antenna based on higher order modes driven by one layer of a substrate-integrated waveguide is demonstrated in [11]. A four-port frequency scanning antenna is presented in [12] with quadri-polarization at 33 GHz. Furthermore, in [13], [14] a dual-band polarization-flexible CRLH antenna is investigated in the ISM band.

As it is mentioned, due to lack of PIN diode availability in millimeter-wave, the phase difference for adjusting the polarization needs to be obtained from an external active circuit. Hence, the antenna itself should add the least complexity to the system. In addition to low complexity, the antenna should be planar to be integrated with other planar circuitries. All of the studies except [12], [13] and [14] require complex feeding network in a multi-layer structure which increases the size, complexity, and weight of the system.

In this letter, a compact single-layer polarization- adjustable antenna is presented at 28 GHz. Four slots are etched on the cavity to generate a hybrid mode. Since the antenna layout is symmetrical, a dual-polarized antenna can be realized and by providing a phase difference between two ports, six types of polarization can be achieved.

## II. Antenna Structure and Working Mechanism

The proposed antenna is illustrated in Fig. 1. A cylindrical cavity is realized on a substrate with post vias, and four ring slots are etched on the cavity as radiating elements. To excite the cavity, a coaxial probe has been used. By applying an offset to the feed point, higher order modes can be stimulated. The antenna is designed on Rogers 5880 substrate with the thickness of 0.787 mm, $\varepsilon_r$=2.2 and loss tangent of $tan\delta$=0.0009. In this study, $TM_{31}$ and $TM_{12}$ are exploited. The cavity radius is chosen to be 8 mm so the cut-off frequency of the aforementioned modes would lie around 28 GHz. Loading a slot on any cavity can alter the cut-off frequency and field distribution of each mode. Here, when the slots are loaded in the cavity, $TM_{31}$ and $TM_{12}$ are shifted to higher frequencies, combine and generate a hybrid mode from 25 GHz to 29.5 GHz.

P. Aghabeyki and G. Moradi are with Applied Electromagnetic Lab, Electrical Engineering Department in Amirkabir University of Technology, Tehran, 15914, IR (e-mail: P_agh@aut.ac.ir, Ghmoradi@aut.ac.ir).
M. Hamidi is with the Department of Electrical and Computer Engineering in Tarbiat Modares University, Tehran, 14115-194 IR (e-mail: Masoud.hamidi@modares.ac.ir).
P. Mousavi is with Intelligent Wireless Technology Group IWT, Electrical Engineering department in University of Alberta, pmousavi@ualberta.ca.

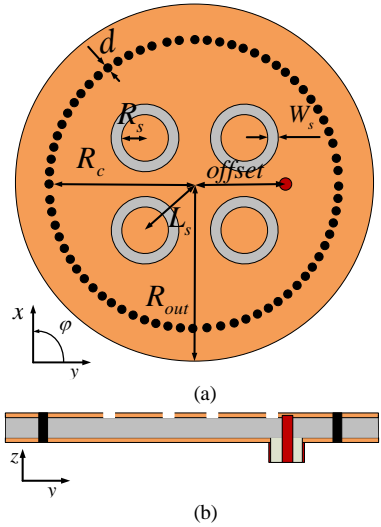

Fig. 1. Antenna configuration (a) top view (b) side view

The hybrid mode is discernable from antenna input resistance as illustrated in Fig. 2. Electric field expressions for $TM_{31}$ and $TM_{12}$ are as follows [15]:

$$E^z_{TM_{31}} = A J_3(\beta_\rho \rho) \times [\sin(3\varphi) + \cos(3\varphi)] \quad (2)$$

$$E^z_{TM_{12}} = A J_1(\beta_\rho \rho) \times [\sin(\varphi) + \cos(\varphi)] \quad (3)$$

where $\beta_\rho = \chi_{mn}/R_c$ and $\chi_{mn}$ represents the *n*th zero (n=1, 2, 3...) of Bessel function $J_m$ of the first kind and of order *m* (m=1, 2, 3...). Electric Field distribution of the hybrid mode is displayed in Fig. 10. The shapes are obtained from CST Microwave Studio and the summation of field expressions presented in (2)-(3).

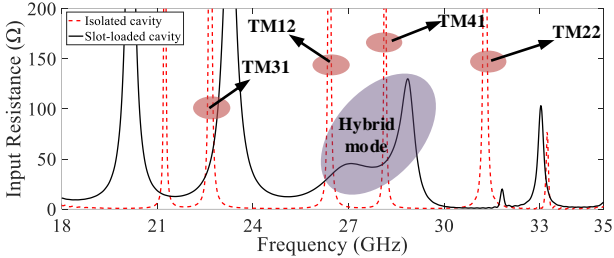

Fig. 2. Input resistance for isolated and slot loaded cavity

A side view of the electric field vectors in the cavity is depicted in Fig. 12(a). The electric field has four peaks in the cavity, with nulls in the center of slots. Electric field vectors distribution on slot aperture is shown in a closer view in Fig. 12(b). Electric fields on all slots are almost in phase in the same direction. Each slot generates two equivalent magnetic monopoles, and overall, we have an array of 8 magnetic monopoles. The alignment of magnetic monopoles yields a high gain.

*A. Dual-Polarized Configuration*

Due to the symmetrical structure, by adding another port with 90° rotation with respect to center of the cavity, one can develop a dual-polarized antenna. In addition, to increase the bandwidth, shorting pins are added to the cavity as shown in Fig. 5. Antenna design parameters are listed in TABLE I.

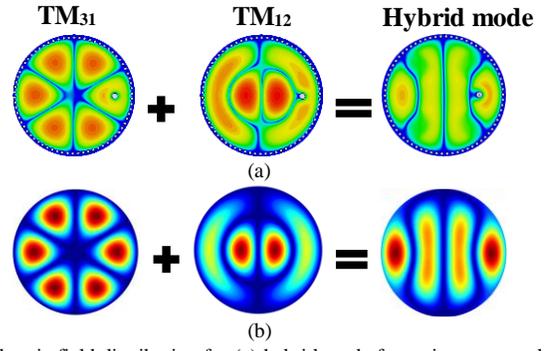

Fig. 3. Electric field distribution for (a) hybrid mode formation extracted from CST and (b) summation of $TM_{31}$ and $TM_{12}$ field expressions.

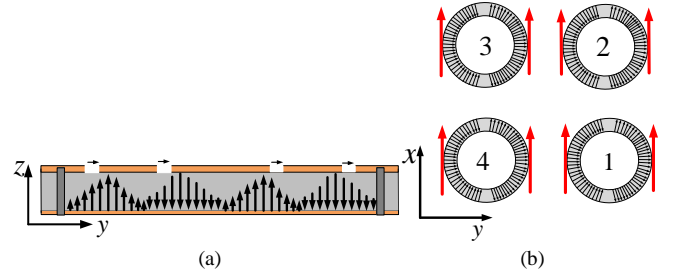

Fig. 4. Electric field vector orientation (a) along the radial direction and (b) on the slot aperture for the hybrid mode.

The effect of adding shorting pins on the reflection coefficient can be observed in Fig. 6. By adding the pins, antenna bandwidth increases and the isolation is improved for higher frequencies but deteriorated for the lower ones. Although the ports are connected directly to the same cavity, the isolation is satisfactory since each port is in the null of the electric field that is excited by its counterpart. The broadband characteristics of the antenna are due to the presence of two modes in close proximity and the stability of the resulting hybrid mode. A parametric study has been carried out on reflection coefficient, isolation, input impedance, and maximum gain. As shown in Fig. 7(a), the increase of $L_S$ would improve the matching over the bandwidth and reduce the isolation. As the value is increased, the peak gain is decreased for the first resonance, and the input resistance increases for the second resonance. Fig 7(b) displays the effect of *offset*. As it shifts toward higher values, the matching is improved over the bandwidth, and the isolation is reduced. Since the probe location moves from the peak of the electric field of one mode ($TM_{31}$) to another ($TM_{12}$), the increase of offset, decreases the first resonance input resistance and increases the second resonance input resistance. The peak gain remains almost unaffected over the bandwidth because there is no change in the dimension and location of the slot. Fig. 7(c) illustrates the impact of $R_S$. As the value increases, matching is improved over the bandwidth and the resonant of the second mode shifts toward lower frequencies whereas the isolation increases. Moreover, the input resistance for the first and second modes is decreased and increased, respectively. Also, the peak gain is enhanced in the bandwidth of interest. Fig. 7(e) demonstrates the effect of $W_S$. As it shifts toward higher values, the matching for the frequencies between the two resonances is enhanced.

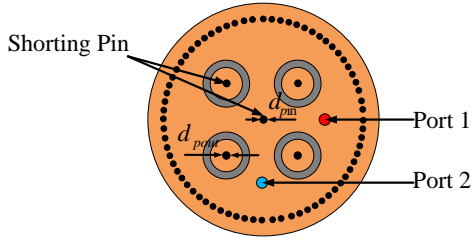

Fig. 5. Dual polarized configuration antenna

TABLE I
ANTENNA DESIGN PARAMETERS

| Para. | Value(mm) | Para. | Value(mm) | Para. | Value(mm) |
|---|---|---|---|---|---|
| $R_S$ | 1.69 | $R_C$ | 8.25 | $d_{pin}$ | 0.7 |
| $R_{out}$ | 11.75 | $W_S$ | 0.45 | $d_{pout}$ | 0.5 |
| offset | 5.35 | $L_S$ | 4.65 | $d$ | 0.5 |

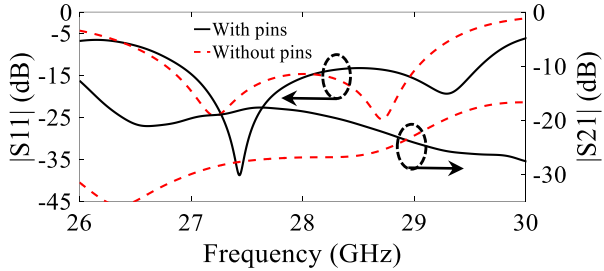

Fig. 6. Reflection coefficient and isolation with and without shorting pins.

Isolation for the first resonance is improved and for the second one is deteriorated. The input resistance is increased for the frequencies between the two modes. In fact, the increase of $W_S$ escalates the loading effect on the cavity thus removes the degeneracy. Also, the peak gain is increased for the first resonance.

*B. Flexible Polarization*

The dual-port configuration enables polarization adjustability with the help of the relative phase difference between the two ports. In the practical sense, the phase difference should be generated via millimeter-wave circuitry, integrated beneath the antenna. For measurement in the Lab, the phase difference can be achieved via commercially available power dividers and couplers. A set of phase and magnitude states are demonstrated in TABLE II for six types of polarization.

### III. EXPERIMENTAL RESULTS AND DISCUSSION

An optimization has been carried out with CST Microwave Studio, and a set of final parameters are obtained. The photograph of the fabricated dual-port antenna is shown in Fig. 8. First, the dual polarization capability of the antenna has been tested to validate the simulation results. S-parameters, maximum gain and radiation efficiency are illustrated in Fig. 9. Since the design is symmetrical, only s-parameters are shown for one port. The port-to-port isolation which is a fundamental factor in a dual-polarized antenna is under -X dB. Antenna $|S_{11}|$ is less than -10 dB from 26.5 GHz to 29.5 GHz which covers the entire Standard n257 5G band. For radiation measurement, one port is connected to the source while the other port terminated in a matched load.

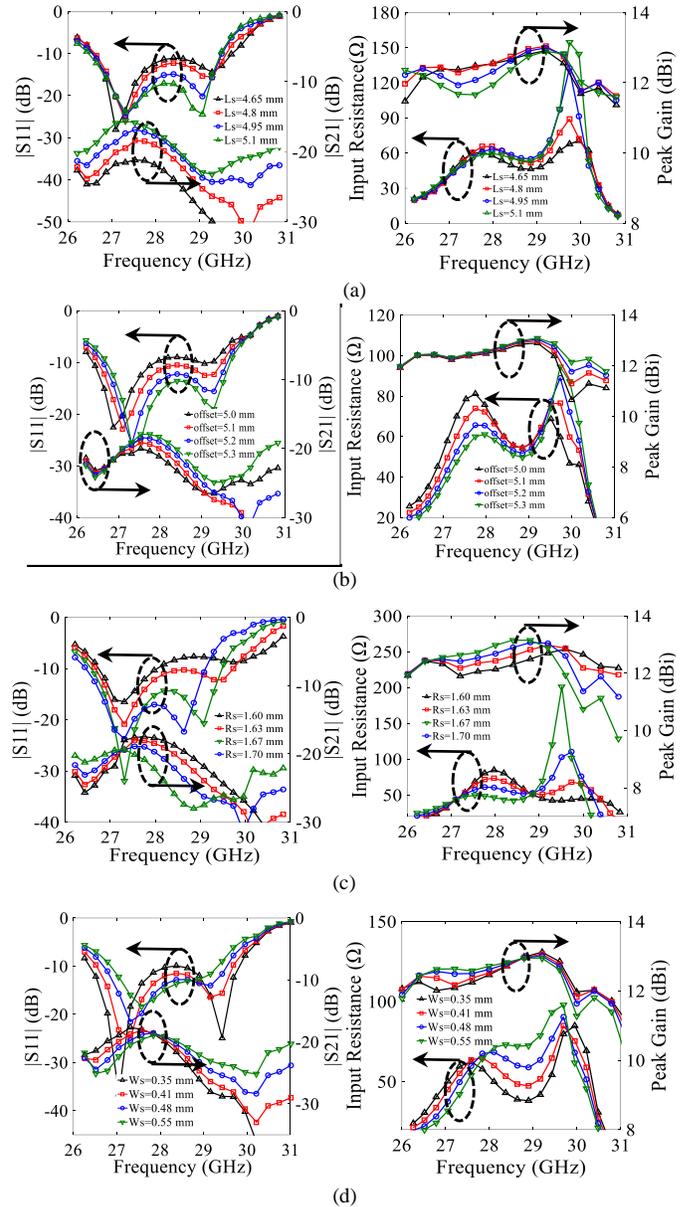

Fig. 7. Simulated $|S_{11}|$, $|S_{21}|$ (left), gain and input resistance (right) for different values of (a) $L_S$, (b) *offset*, (c) $R_S$ and (d) $W_S$

The gain is over X dBi with a maximum of X dBi at 29 GHz. Measured radiation efficiency is over X% in the entire bandwidth. The radiation pattern for horizontal polarization in E-plane and H-plane at 28 GHz is depicted in Fig. 10. The cross-polarization level is X dB less than the co-polarization level in the boresight direction.

For measuring the slant polarization, since the configuration is symmetrical, only $\varphi = 45°$ is considered. The input ports of the antenna is connected to a power divider that renders two outputs with equal magnitude and 180° phase difference to generate a slant polarization. Simulated and measured gain and efficiency is shown in Fig. 11. The maximum gain is above X dBi over the bandwidth. Antenna radiation pattern is plotted in Fig. 12. The cross-polarization level is -X dB in the broadside direction.

TABLE I
AMPLITUDE AND PHASE FOR SIX TYPES OF POLARIZATION

| Polarization mode | Linear along y-direction | Linear along x-direction | Linear along $\varphi = 45°$ | Linear along $\varphi = 135°$ | Right Hand Circular Polarization (RHCP) | Left Hand Circular Polarization (LHCP) |
|---|---|---|---|---|---|---|
| Amplitude and phase state | $A_2 = 0, A_1 \neq 0$ | $A_1 = 0, A_2 \neq 1$ | $A_1 = A_2$ $\|\varphi_1 - \varphi_2\| = 180°$ | $A_1 = A_2$ $\varphi_1 = \varphi_2$ | $A_1 = A_2$ $\varphi_1 - \varphi_2 = -90°$ | $A_1 = A_2$ $\varphi_1 - \varphi_2 = +90°$ |
| Total E-field on the aperture | 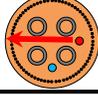 | 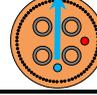 | 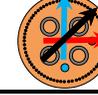 | 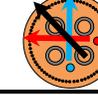 | 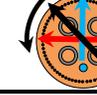 | 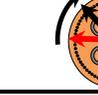 |

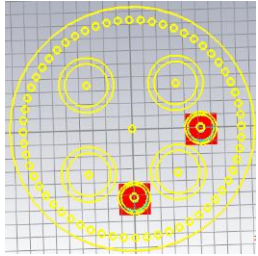

Fig. 8. Photograph of the fabricated antenna

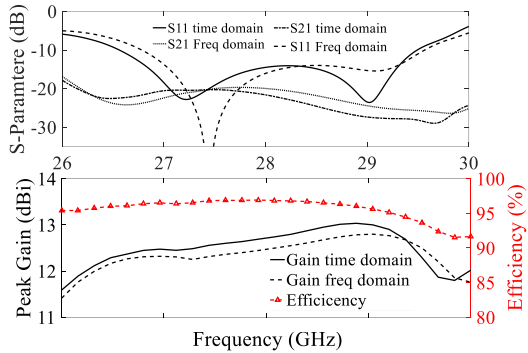

Fig. 9. Simulated and measured $|S_{11}|$, $|S_{21}|$, gain and efficiency for dual-polarized antenna

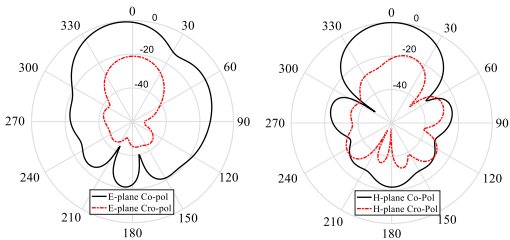

Fig. 10. Dual-polarized antenna radiation pattern at 28 GHz

To evaluate the circular polarization performance, a hybrid coupler with equal magnitude output and 90° phase difference is required to generate circular polarization. Axial ratio (AR) and maximum gain are depicted in Fig. 11. The 3-dB AR bandwidth covers the entire impedance bandwidth, and the maximum gain is over X dBi. The antenna radiation pattern for RHCP is displayed in Fig. 14. The cross-polarization is –X dB in the boresight. Due to the symmetrical layout, other polarizations (vertical, $\varphi = 135°$ and LHCP) indicate the same results.

## IV. CONCLUSION

A dual-port polarization-adjustable antenna is realized by introducing four ring slots on a cylindrical SIW cavity. The direct probe feed excitation eliminates the feeding network.

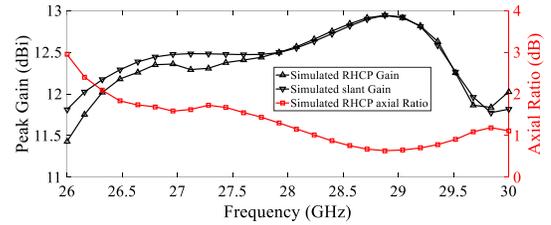

Fig. 11. Peak gain for slant and RHCP polarization and axial ratio at 28 GH

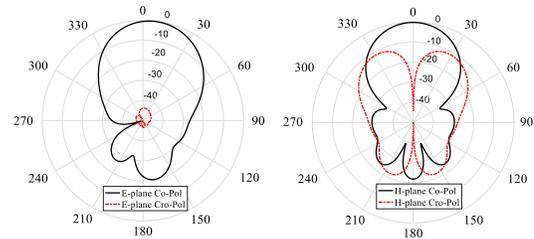

Fig. 12. Slant polarization radiation pattern at 28 GHz

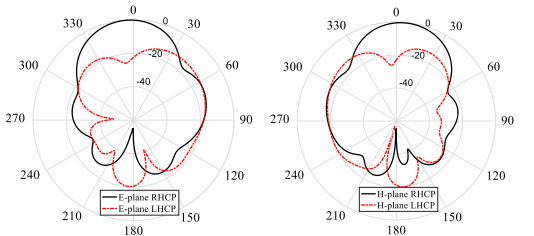

Fig. 13. RHCP polarization radiation pattern at 28 GHz

Owing to the symmetrical configuration, any given polarization can be achieved via proper amplitude and phase difference between the two ports. In dual-polarized mode, the antenna offers –X dB port-to-port isolation and X % fractional bandwidth in 28 GHz. For flexible polarization performance, all six types of major polarizations can be obtained with a high gain and good cross-polarization level. The antenna features low-profile, low cost with an adequate bandwidth and gain for millimeter-wave applications. Furthermore, the absence of feeding network adds the least complexity to a communication system.